# Role of electronic thermal transport in amorphous metal recrystallization: a molecular dynamics study

Zachary D. McClure[1], Samuel Temple Reeve[1], and Alejandro Strachan[1*]

[1]School of Materials Engineering and Birck Nanotechnology Center
 Purdue University, West Lafayette, Indiana 47906 USA.

## Abstract

Recrystallization of glasses is important in a wide range of applications including electronics and reactive materials. Molecular dynamics (MD) has been used to provide an atomic picture of this process, but prior work has neglected the thermal transport role of electrons, the dominant thermal carrier in metallic systems. We characterize the role of electronic thermal conductivity on the velocity of recrystallization in Ni using MD coupled to a continuum description of electronic thermal transport via a two-temperature model. Our simulations show that for strong enough coupling between electrons and ions, the increased thermal conductivity removes the heat from the exothermic recrystallization process more efficiently, leading to a lower effective temperature at the recrystallization front and, consequently, lower propagation velocity. We characterize how electron-phonon coupling strength and system size affects front propagation velocity. Interestingly, we find that initial recrystallization velocity increases with decreasing in system size due to higher overall temperatures. Overall, we show that a more accurate description of thermal transport due to the incorporation of electrons results in better agreement with experiments.

## 1. Introduction

Amorphous materials are attractive for a wide range of applications from sports equipment[1], structural support[2,3], functional bio-replacements[4], phase-change memory (PCM)[5], magnetic devices[6], and reactive composites[7,8]. In many of these applications, the mechanisms and kinetics associated with recrystallization dominate performance. PCM devices operate by reversible switching between an amorphous phase of high resistance and a crystalline phase of low resistance and have attracted significant interest for scalable non-volatile memory applications[5,9]. In these devices the speed of recrystallization from the glassy phase in PCMs is a critical performance metric associated with switching speed. The degree of undercooling, thermal history, and material stability are known to affect recrystallization behavior and this knowledge has been used to improve the operation of PCM devices[10]. Amorphous components have also found application in reactive alloys, where a higher free energy and driving forces than their crystalline counterparts improve performance[8] for potential use in lead-free explosive primers[11,12], and soldering[13,14]. In this case, the stability of the amorphous phase is important.

Amorphous metals are thermodynamically metastable and can be synthesized through multiple routes[15], including fast cooling from the liquid[16], or through aqueous chemical reduction[17,18]. Given appropriate thermal or mechanical stimuli these glasses can recrystallize via a solid-to-solid, exothermic phase transition[19]. As the material recrystallizes, the released energy creates a self-propagating reaction front. The velocity of this reaction front is dependent on intrinsic material properties including exothermicity of the reaction, thermal conductivity of the material, microstructure, and defects. Despite the significant interest and importance of glassy metals several questions remain unanswered. For example, large scale predictive atomistic simulations have been able to characterize recrystallization behavior in amorphous materials[8]; however, these methods ignore important processes in electron dominated thermal transport.

Self-propagating recrystallization have been observed experimentally in, for example, amorphous silicon[20,21] and amorphous Ni-Fe foils[22]. Molecular dynamics has contributed to our understanding of recrystallization in semiconductors and insulators, where phonon transport dominates the thermal properties[23,24]. MD simulations of recrystallization in metals have been explored to a lesser degree. However, Manukyan et al. studied self-propagation velocities in agglomerates of Ni nanoparticles and compared to experiments[8]. The authors found that the simulations predicted significantly higher propagation velocities. While several effects contribute to this discrepancy, we hypothesize that the underestimation of thermal transport in metals through missing electronic effects is a critical factor. In this paper, we explore the effect of thermal conduction by electrons in recrystallization of amorphous Ni.

Carriers for thermal transport in metals are phonons (or the equivalent collective ionic modes in non-crystalline solids) and conduction electrons; while they have a smaller specific heat the latter dominate thermal transport[25]. Standard MD simulations describe thermal transport through ionic processes, but do not include electronic contributions. Multiple methods have been developed that can account for electron interaction and transport including the two-temperature model (TTM), which models electron thermal transport through a diffusive heat equation[26,27], as well as dynamics with implicit degrees of freedom (DID), where electrons are modeled as degrees of freedom (DoF) for each atom[28]. Here, we add electronic effects with the TTM to simulate more realistic behavior for amorphous metal recrystallization.

The remainder of the paper is organized as follows: Section 2 includes amorphous sample preparation, an introduction to the TTM, and verification of TTM inputs. Section 3 shows results for recrystallization without electronic effects, and recrystallization simulations varying the electronic inputs in Section 4. Finally, we discuss our results and conclude in Section 5.

## 2. Simulation Methods

### 2.1 Atomistic model of Ni and sample preparation

All MD simulations were performed using the LAMMPS (Large-scale Atomic/Molecular Massively Parallel Simulator) software package developed by Sandia National Laboratory[29]. Visualization was performed using OVITO (Open Visualization Tool)[30] with the polyhedral template matching (PTM) algorithm[31] to identify atoms belonging to crystalline and amorphous phases. Throughout, atoms with an FCC local environment are colored green and atoms with unidentified coordination are white. An embedded-atom method (EAM) potential for Ni was used in our simulations, parameterized by Mishin[32] (accessed through the NIST Interatomic Potentials Repository[33] under Ni-Al 2009). This model was parameterized using density functional theory (DFT) formation energies and experimental results including cohesive energy, lattice parameter, and elastic constants. To assess the accuracy of the interatomic potential to describe amorphous Ni configurations we computed the melting temperature using the coexistence simulation technique[34] and heat of fusion as the enthalpy difference between liquid and crystalline samples at the melting temperature. The melting temperature predicted by the potential is 1750 K and the predicted heat of fusion is 0.19 eV/atom, both in good agreement with the experimental values of 1728 K and 0.181 eV/atom, respectively[35].

## 2.2 Sample preparation

All recrystallization simulations were performed with a crystalline seed in contact with an amorphous sample, extended in one direction. A similar simulation was previously performed to investigate crystallization of amorphous Ni nanoparticles[8]. A timestep of 1 fs was used throughout, with damping constants of 0.1 ps and 1.0 ps for the Nose-Hoover thermostat and barostat, respectively. All simulations were run at 1 atm pressure.

Crystalline seeds were generated by replicating the FCC unit cell 5x10x10 times. The resulting crystalline seed initially measured 1.76 nm by 3.52 nm by 3.52 nm with periodic boundary conditions and was equilibrated using the isothermal-isobaric (NPT) ensemble at 300K for 100 ps. Amorphous samples of different lengths were generated via melting, deformation, and quenching. The initial configurations were obtained by replicating the FCC unit cell 10x10 along the directions of the crystal/amorphous interface leading to cross-sectional dimensions of 3.52 nm by 3.52 nm with periodic boundary conditions. The unit cell was replicated between 100 and 300 times along the recrystallization direction resulting in initial system lengths between 35.2 and 105.6 nm. A temperature ramp from 300 to 3000K in 100 ps was used to melt the sample under NPT conditions. To prevent strain in the final structure, the system was deformed in the melt at 3000K to match the transverse directions of the crystalline seed lattice, conserving the total volume of the cell. The system was then quenched from 3000 to 300K in 100 ps under NPT conditions. The quenched amorphous samples were equilibrated at 300K with fixed cross section and NPT conditions along the recrystallization direction. Five independent samples for each system length were taken during equilibration in increments of 10 ps.

Samples for recrystallization were generated by combining the crystalline seed and amorphous bulk. The two systems were added to the same simulation cell (already possessing identical transverse dimensions) with a gap of 5 Å, see Figure 1. A vacuum layer of 3.5 nm was added in the longitudinal direction to ensure only one crystal/amorphous interface. The total system was then relaxed with energy minimization and equilibrated for 5 ps under NPT at 300K.

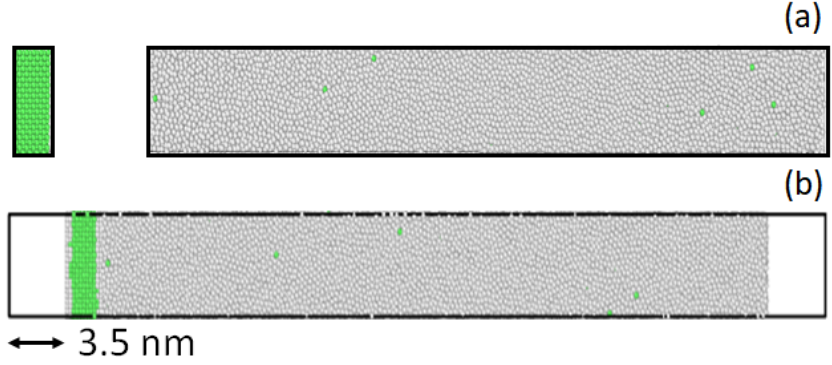

Figure 1: (a) Crystalline seed and amorphous bulk with periodic boundary conditions after separate equilibration. (b) Combined structure for recrystallization experiments with added vacuum layers.

## 2.3 Recrystallization simulations

The approach for investigating amorphous nickel recrystallization was first tested without electronic effects. Two different methods were applied to investigate the amorphous nickel recrystallization. The first was under isothermal and isobaric conditions (NPT ensemble) throughout the recrystallization. The amorphous and seed regions were heated to various set temperatures, and then held to recrystallize for at least 0.5 ns. The second method simulated an adiabatic reaction by exposing the seed to a fast thermal impulse while maintaining the amorphous region at 300K. The set temperature of the seed was ramped to various values in 8 ps and held for 2 ps. The subsequent process can be described as an adiabatic MD simulation, using an isobaric-isenthalpic (NPH) ensemble.

## 2.4 Two-temperature model MD

The TTM used here solves the equations of motion for atoms using MD, adding an electronic temperature field defined over a grid overlapping with the atomistic system. Electrons and atoms interact and exchange energy which affects the ionic equations of motion via an additional friction coefficient:

$$m \frac{\partial v_i}{\partial t} = F_i(t) - \gamma_i v_i + \tilde{F}(t) \qquad (1)$$

where v, m, and F are the velocity, mass, and standard MD force (gradient of potential energy) of atom i, respectively. The term including friction coefficient $\gamma_i$ is the energy loss due to electron-ion interactions and the stochastic force term, $\tilde{F}(t)$, is determined by the local electron temperature as a Langevin thermostat.

$$\tilde{F}(t) = -\gamma_L v(t) + \xi(t) \tag{2}$$

The stochastic force is determined by a Langevin friction coefficient of the system, $\gamma_L$, and a random force obtained from a Gaussian probability distribution, $\xi$, giving the effect of background noise.

Energy loss between electrons and phonons occurs via two pathways depending on the velocity of an atom[36,37]. For high atomic velocities (e.g. approximated as 5.4 km/s for Fe[27]) the valence electrons slow the atoms through a process known as electronic stopping. At lower velocities, of interest in this work, electron-ion interactions act to bring the two sub-systems to thermal equilibrium. Atomic motion is slowed down or sped up proportionally to the difference in the electronic and atomic temperature. $\gamma_s$ is the friction coefficient due to electronic stopping and $\gamma_p$ is the friction coefficient due to electron-ion interactions. At relatively low atomic temperatures only the electron-phonon interaction plays a role and $\gamma_s$ is accordingly set to zero. For an atomic velocity $v_i$, the interaction between electrons and phonons will change above or below a cutoff velocity $v_0$.

$$\gamma_i = \gamma_s + \gamma_p \quad for \quad v_i > v_o \tag{3}$$

$$\gamma_i = \gamma_p \quad for \quad v_i \leq v_o \tag{4}$$

The electronic degrees of freedom transport heat and the TTM solves the heat diffusion equation for the electronic temperature at discrete grid points throughout the atomic structure. Our electron grid was set with one point for each two lattice spacing units in each direction and was found to be well converged. At each grid point the heat diffusion equation is coupled between the atoms and the electrons:

$$C_e \rho_e \frac{\partial T_e}{\partial t} = \nabla(\kappa_e \nabla T_e) - g_p(T_e - T_a) + g_s T_a' \tag{5}$$

where $C_e$ represents the electronic heat capacity, $\rho_e$ the electronic grid density, and $\kappa_e$ the electronic thermal conductivity. For our relatively low temperature simulations we ignore friction effects from electronic stopping by setting $\gamma_s$ to zero. In turn, the coupling parameter for electron-stopping, $g_s$, is zero and the electron-phonon coupling parameter, $g_p$, is:

$$g_p = \frac{3Nk_b \gamma_p}{\Delta V m} \tag{6}$$

where N is the total number of atoms in the electronic grid, $k_b$ is the Boltzmann constant, $\Delta V$ is the electronic grid volume, and m is the atomic mass. The related inverse relaxation time, or the coupling coefficient, $\chi$, is:

$$\chi = \frac{\gamma_p}{m} \qquad (7)$$

The TTM implementation used was extended to allow vacuum layers within laser ablation simulations[38,39]. This was important for the simulation to ensure recrystallization along a single front (see Figure 1). Finally, we note that electronic properties depend on the local atomic structure. For example, the thermal conductivity would be reduced in the amorphous regions. In this first study, we ignore such effects.

## 2.5 TTM verification tests and input parameters

A series of TTM MD simulations under adiabatic conditions (NVE ensemble) were performed for verification purposes and to test input parameters. A system of 4,000 atoms (40x5x5 unit cells) was created and equilibrated at 300K using NPT. A series of TTM simulations were carried out to assess the effect of electronic specific heat, Figure 2(a), and electron-phonon coupling constant, Figure 2(b), on the equilibration of the electronic and ionic temperatures by setting the initial temperature of the electronic subsystem to 600K.

**Role of electronic specific heat.** Figure 2(a) shows the time evolution of electronic and ionic temperatures for simulations with three different electronic heat capacities: 0.3, 3, and 3 $k_B$ per atom. All three simulations use a coupling constant of $\gamma_p$ = 10 g mol$^{-1}$ ps$^{-1}$, which results in a coupling coefficient of $\chi$ = 0.17 ps$^{-1}$. The results of the simulations are as expected. Recall that within the classical harmonic approximation, the heat capacity of a solid in 3D is 3$k_b$ per atom. By setting the electronic specific heat to an unrealistically high value of 3$k_b$ per atom both sub-systems equilibrate at a temperature approximately half way between their initial values. However, an electronic heat capacity one-tenth of the atomic subsystem shows a more realistic process where electrons equilibrate to a temperature close to the initial ionic temperature[28,40]. The heat capacity of electrons is temperature dependent, but to simplify the numerical aspects of our simulations we use a constant value 0.3 $k_b$ per atom for the remainder of this paper. This is a reasonable approximation for our problem of interest since the electronic heat capacity of Ni varies from 0.29 to 0.48 $k_b$ per atom in the temperature range 300-1500 K[40].

**Electron-phonon coupling.** Electron phonon coupling constants can be determined experimentally[41,42], as well as from first principles[40]. The electron-phonon coupling constant for Ni has been reported with a range of $\chi = 0.10 – 0.29$ ps$^{-1}$ from DFT simulations[43]. Estimates of the electron-phonon coupling coefficient through numerical analysis resulted in $\chi = 1.0$ ps$^{-1}$ for Ni and 0.05, 0.03, and 1.5 ps$^{-1}$ for Ag, Cu, and Fe, respectively[44]. Figure 2(b) characterizes the process of equilibration between electrons and ions with coupling constants varying by three orders of magnitude and a fixed heat capacity of $C_e = 0.3$ k$_b$ per atom. The results show the expected behavior for comparable values to those above and demonstrate that for Ni, equilibration occurs with a timescale of approximately one ps.

**Electron thermal conductivity.** Another important parameter in the TTM is the electronic thermal conductivity, given by the product of the electron grid density $\rho_e$, the electron heat capacity $C_e$, and the electron thermal diffusivity $D_e$. For the TTM simulations $D_e = 2$ cm$^2$ s$^{-1}$ was used[39]. This corresponds to an electron thermal conductivity for Ni of ~74 W m$^{-1}$ K$^{-1}$. Electron thermal conductivity accounts for nearly 90% of transport in bulk and at metal interfaces[25]. The experimental thermal conductivity for Ni is 90 W m$^{-1}$ K$^{-1}$ [45] and is the sum of both electronic and phonon thermal contributions. MD calculations of the phonon contribution range from 2-10 W m$^{-1}$ K$^{-1}$ [46]. Using the sum of the calculated electronic and phonon thermal conductivities our input for the TTM matches experiments well.

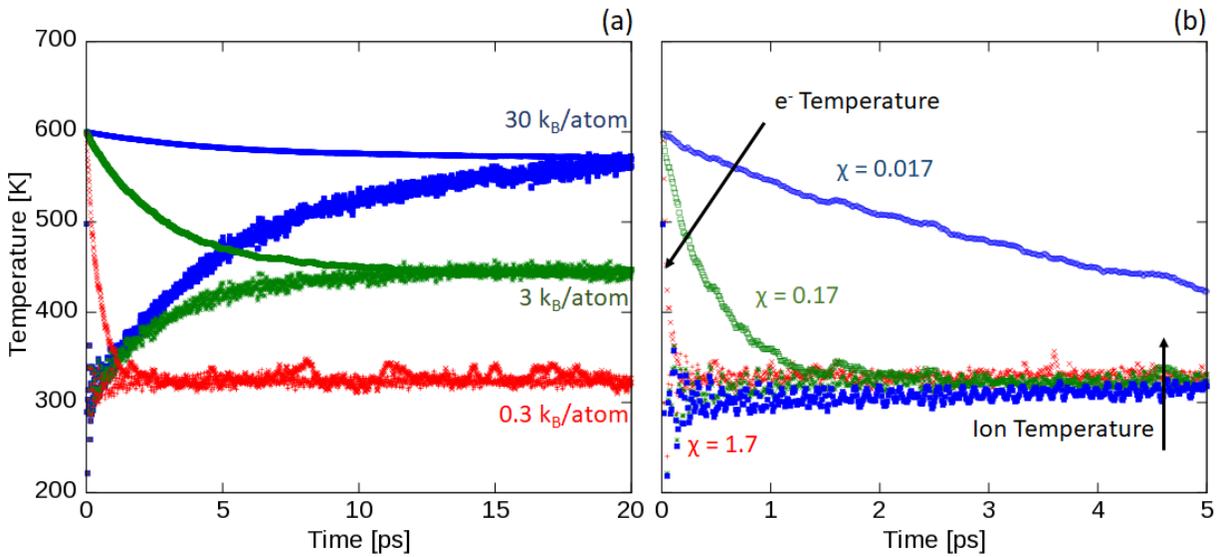

Figure 2: Temperature equilibration between atomic (T$_0$=300K) and electronic (T$_0$=600K) subsystems. (a) Varying electronic heat capacity with $\chi = 0.17$ ps$^{-1}$. (b) Varying electron-ion coupling constant with $C_e = 0.3$ k$_b$/atom.

## 3. Recrystallization without electronic effects

We begin with standard MD recrystallization, without added electronic effects and under NPT conditions. Figure 3 shows the atomistic structure during recrystallization for a representative system at 1000K. The crystalline seed acts as a heterogeneous nucleation site for the amorphous metal and the recrystallization front moves across the system.

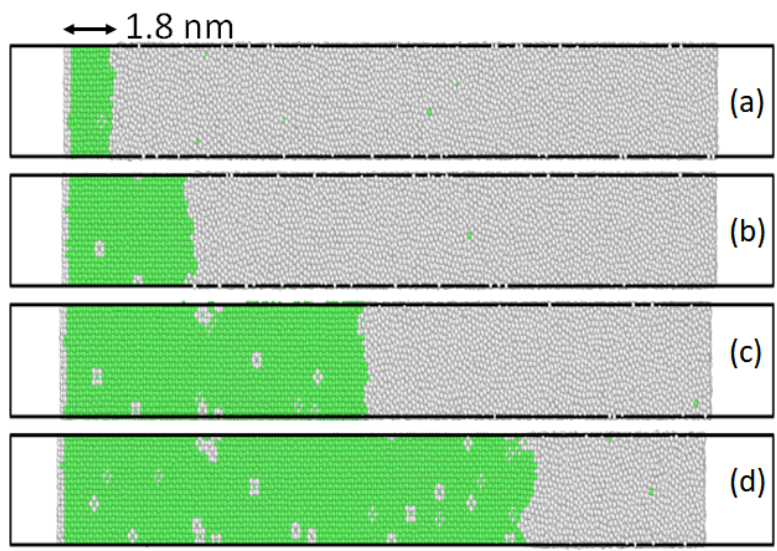

Figure 3: Snapshots of MD simulation of recrystallization of amorphous Ni at 1000K at increasing time (a) 10, (b) 100, (c) 300, (d) 500 ps. Atoms in green are FCC coordination, while white atoms are unidentified.

Under isothermal conditions, the crystallization velocity vs. temperature exhibits a maximum at a value below the melting temperature that compromises between driving force for crystallization and kinetics. At lower temperatures kinetics dominate and reduce crystallization front speed. At higher temperatures the thermodynamic driving force for crystallization (the free energy difference between the amorphous and crystalline phases) decreases. Figure 4 shows the average recrystallization front velocity as a function of temperature (black circles). Throughout all simulations, the velocity is calculated from the slope of the linear recrystallization front position as a function of time, as shown in Figure S1 in the supplementary material. The velocity increases with increasing temperature up to 1200 K. Beyond 1200 K, a sharp drop in velocity is observed; this is consistent with previous MD simulations of Ni recrystallization[8]. Isothermal conditions are artificial as they remove or add energy to the system in non-physical ways. In addition, a global thermostat couples with the total kinetic energy of the system and is unaware of local temperature variations, such as the local heating around the crystallization front. Adiabatic conditions represent experimental conditions for fast recrystallization processes more accurately. Thus, constant pressure and enthalpy (NPH ensemble) simulations were conducted and red squares in Figure 4 show the resulting front velocity as a function of the initial seed temperature. As expected, we observe a weak dependence of recrystallization velocity on initial seed temperature as the system uses the heat generated from recrystallization to continue the process and evolves towards a steady state. This behavior is the focus for the remainder of this work.

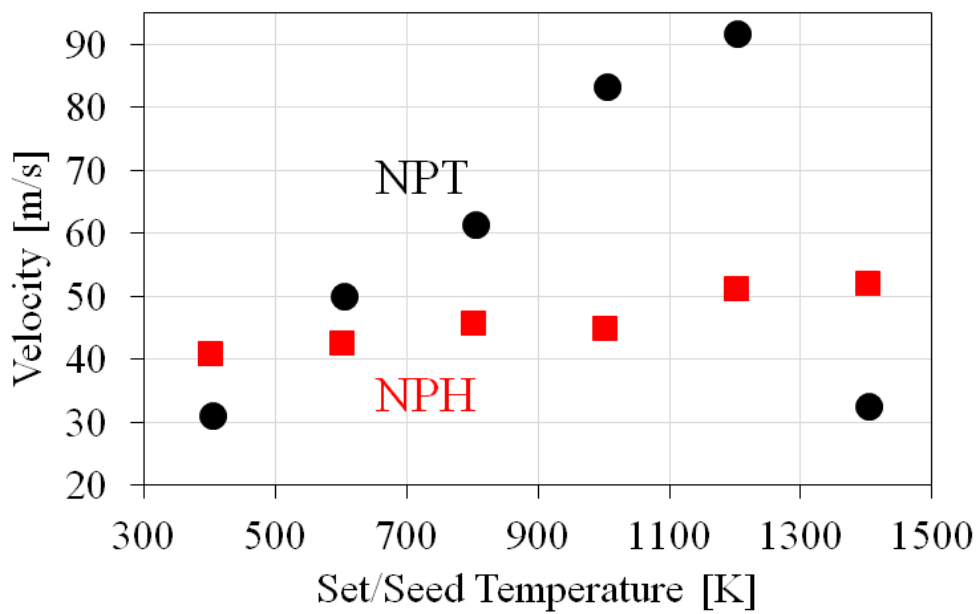

Figure 4: Crystallization velocities for isothermal-isobaric (NPT) and adiabatic (NPH) simulations. Temperatures shown represent the set temperature for the entire system in NPT, and the initial seed temperature in NPH simulations.

The details of the process under adiabatic conditions can be seen in Figure 5(a) where we show temperature profiles at various times. At 10 ps, the crystalline seed has been heated to its set point temperature while the amorphous region was held at 300K. By 100ps, the temperature of the crystallized region has dropped to 600K, as some heat has diffused away from the recrystallization front to the rest of the amorphous sample. Following this initial stage, we observe steady state propagation of the recrystallization front where the vertical lines in Figure 5 indicate the location of the front at each time. Throughout the simulation the hottest part of the sample is within the recrystallized region. Heat builds up at the reaction front and only slightly increases the temperature of the amorphous region. Without a sufficient thermal conductivity to carry it away a large amount of thermal energy is provided for recrystallization.

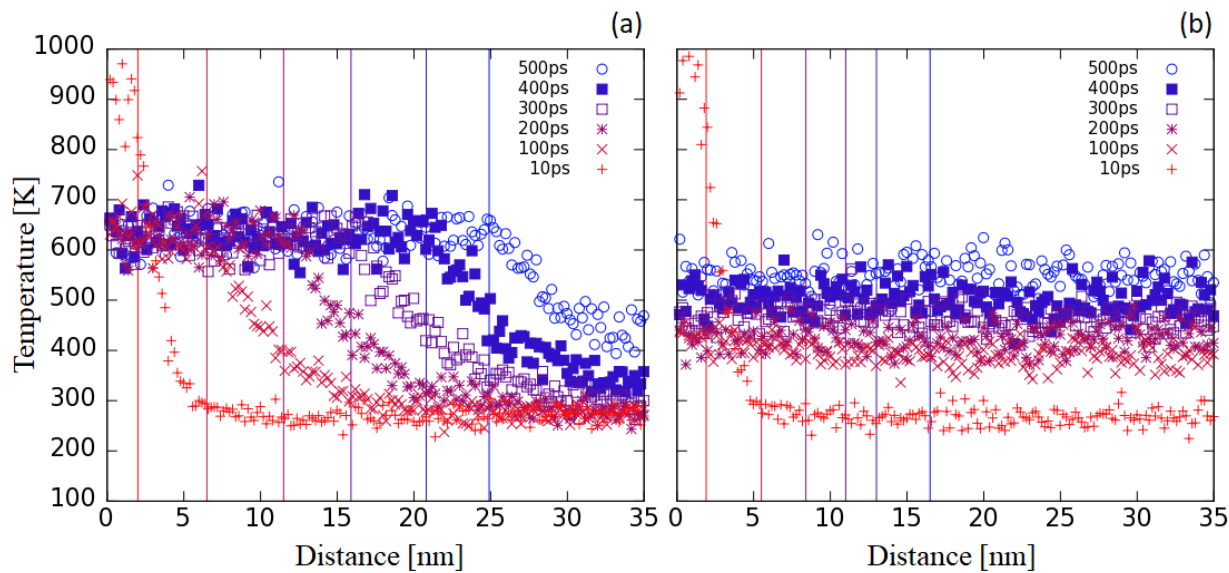

Figure 5: Local temperature along the recrystallization direction for (a) standard MD recrystallization and (b) TTM MD recrystallization with $\chi = 0.17$ ps$^{-1}$. Colored vertical lines indicate recrystallization front position for each time.

## 4. Role of electronic degrees of freedom on recrystallization

Adding the thermal role of electrons using the TTM, with parameters for Ni discussed in Section 2, to the adiabatic recrystallization simulations results in the temperature profiles shown in Figure 5(b). The initial localized temperature in the crystalline seed dissipates through the material in under 100 ps. Instead of a stepped temperature profile that moves with the recrystallization front, the system exhibits a nearly homogenous temperature profile. As the reaction proceeds, the exothermic reactions increase the overall temperature of the system. To better understand the effect of electronic thermal transport we study how the coupling coefficient, $\chi$, and the electronic thermal conductivity, $\kappa_e$, affect the recrystallization process and front velocity.

**Role of coupling strength**. The effect of electron-phonon coupling on front recrystallization velocity is shown in Figure 6(a). For low coupling rates, the energy of recrystallization does not couple to the electronic DoFs sufficiently fast to affect the process. We observe propagation velocities, Figure 6(a) and temperature profiles, Figure 7(a), similar to those corresponding the standard MD simulations without TTM, but we do note a shallower temperature gradient. As the electron-phonon coupling rate is increased, the recrystallization velocity smoothly decreases and then saturates. Once the coupling is strong enough to maintain an equal temperature associated with the two sets of degrees of freedom, an increase in coupling has no effect, leading to the converged velocity. Figure 7(b) shows the effect of electron transport on the temperature profile for a higher coupling, showing earlier times from Fig. 5(b). Notably, as electronic transport plays a role in recrystallization, front velocities become length dependent, shown with two simulation cell lengths in Figure 6(a). As the heat is quickly diffused and spread to more atoms, the larger system will be in average colder, and the overall recrystallization velocity reduced.

Although the temperature profiles and reaction velocities are notably different with varied electronic coupling, the atomic features of the reaction front remain similar. As shown in Figure S2 in the supplementary material, we observe no noticeable difference in the thickness and shape of the reaction front between simulations without electronic coupling and different levels of TTM coupling. In all structures, we find a relatively sharp transition between phases with roughness on the order of 1 nm or a few atomic layers.

**Role of electronic thermal diffusivity**. Similarly, a sweep of electron thermal conductivity was performed, shown in Figure 6(b). As with the coupling parameter, we find two regimes in the electronic thermal conductivity with a smooth transition in between. For small electronic thermal diffusivities (orders of magnitude smaller than the typical values) electronic degrees of freedom have no effect on recrystallization velocity, as phonons are still more effective at distributing heat. Once the electron thermal conductivity is on the same order of magnitude as the phonon thermal conductivity, about 1 W m$^{-1}$ K$^{-1}$, there is a large reduction in recrystallization rate. At saturation the increased electronic conductivity cannot be used effectively by the system as the energy is distributed throughout the entire sample.

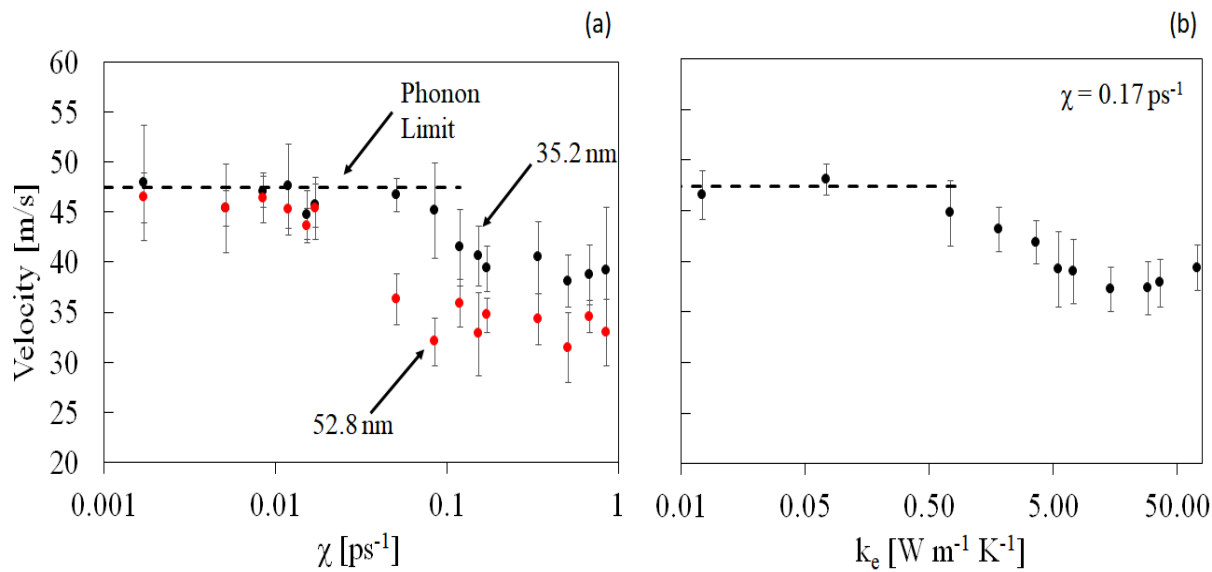

Figure 6: Recrystallization velocity as a function of (a) coupling coefficient $\chi$ for multiple lengths (Black = 35.2 nm, Red = 52.8 nm) and (b) electron thermal conductivity. Phonon limit (standard MD) represented by dashed line.

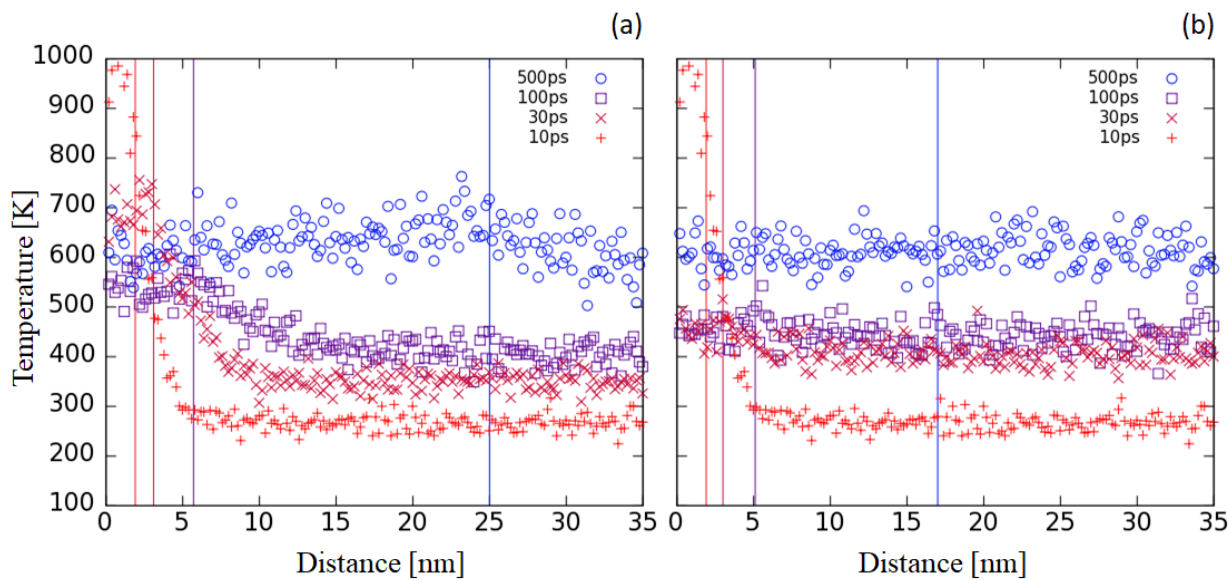

Figure 7: Local temperature along the recrystallization direction for (a) $\chi = 0.017$ ps$^{-1}$ and (b) $\chi = 0.17$ ps$^{-1}$. Colored vertical lines indicate recrystallization front position for each time.

**Size effects**. Since electrons can spread the heat generated over the entire sample, one can expect size effects on the recrystallization velocity. To characterize such effects, a range of sample lengths were generated and recrystallized. Figure 8 compares simulation results with and without electronic degrees of freedom to previously reported experimental values. As noted before, the standard MD with phonon dominated thermal transport has no length dependence on recrystallization velocity due to its low thermal transport. As expected, increasing the sample size for TTM MD simulations results in a reduction of the crystallization velocity due to the reduction in overall temperature. Figure 8 enables an estimation of the crystallization velocity for bulk samples including the role of electrons. The resulting value of 23 m/s is closer to an experimental value of 10-20 m/s for amorphous Fe-Ni foils[22], and 2-17 m/s for amorphous Si[20,47]. It should be noted that additional mechanisms such as interatomic diffusion in the Fe-Ni foils will contribute to recrystallization velocities. However, the alloy system provides a reference point for expected kinetics of metallic recrystallization. We note that experimental recrystallization of amorphous Ni powders has been reported at 0.3 mm/s, significantly lower than our predictions and the other experimental data. However, these Ni samples were agglomerates of nanoparticles with large porosity (~80%) and oxide layers[8], a nanostructure very different from our current MD model.

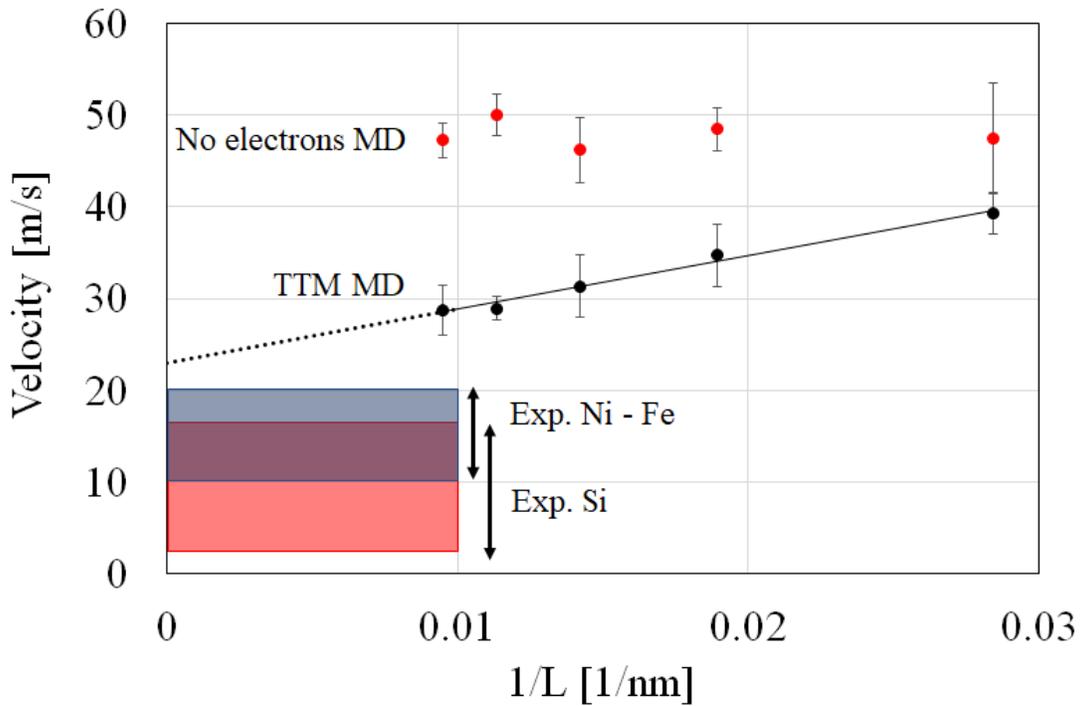

Figure 8: Recrystallization velocity extrapolated to bulk sample lengths from MD without electrons and TTM MD, compared to experimental data.

## 5. Conclusions

In this paper we characterized the role of the thermal transport by electrons on the recrystallization of amorphous Ni. This is done using a two-temperature model coupled to molecular dynamics simulations. The simulation setup uses a heated crystalline seed to start the process of adiabatic recrystallization. We find that the increased thermal conductivity, resulting from the incorporation of electronic degrees of freedom, reduces the propagation velocity of the recrystallization front provided the coupling between ionic and electronic degrees of freedom is fast enough. For weak electron-ion couplings, the recrystallization velocity is not affected by the electronic degrees of freedom as heat from the exothermic recrystallization process is not transferred to the electrons within a characteristic timescale. For coupling coefficients between 0.01 and 1 $ps^{-1}$ electrons can dissipate enough heat from the reaction zone to reduce the effective temperature of recrystallization and, consequently, velocity. When electronic effects are added to thermal transport we find significant length effects on recrystallization velocity as electrons diffuse heat rapidly through the system resulting in a homogeneous temperature profile. Previous estimates of recrystallization in amorphous Ni from standard MD greatly overestimated the recrystallization kinetics; adding electronic contributions to recrystallization results in an estimation of a bulk recrystallization velocity for Ni of 23 m/s.

## Acknowledgements

This work was supported by the Sandia National Laboratory Directed Research and Development program. Sandia National Laboratories is a multimission laboratory managed and operated by National Technology and Engineering Solutions of Sandia, LLC., a wholly owned subsidiary of Honeywell International, Inc., for the U.S. Department of Energy's National Nuclear Security Administration under contract DE-NA0003525. Computational resources from nanoHUB and Purdue University are gratefully acknowledged.